\documentclass{appolb}
\usepackage{graphicx}

\begin{document}
\title{Recent results from NA61/SHINE
\thanks{Presented at Excited QCD 2017}%
}
\author{Maja Ma\'{c}kowiak-Paw{\l}owska for the NA61/SHINE Collaboration
\address{Faculty of Physics, Warsaw University of Technology}
}
\maketitle
\begin{abstract}
NA61/SHINE at the CERN SPS is a fixed-target experiment pursuing a rich physics program including measurements for heavy ion, neutrino and cosmic ray physics. The main goal of the strong interactions program is to study the properties of the onset of deconfinement and to search for the signatures of the critical point.

In this contribution the latest NA61/SHINE results on particle spectra as well as on fluctuations and correlations from p+p, Be+Be, and Ar+Sc energy scans will be presented. The NA61 measurements
will be compared with world data and with model predictions.
\end{abstract}
\PACS{25.75.-q, 25.75.Ag, 25.75.Gz, 25.75.Dw, 25.75.Nq}
  
\section{Introduction}
The NA61/SHINE experiment~\cite{Antoniou:2006mh, Abgrall:2014fa} performs a two dimensional scan in temperature, $T$, and baryo-chemical potential $\mu_{B}$ of the phase diagram of strongly interacting matter by varying system size and energy of the interaction. The program is motivated by the discovery of the onset of deconfinement in Pb+Pb collisions at 30$A$ GeV/$c$ by the NA49 experiment~\cite{Alt:2007aa, Afanasiev:2002mx}, and possible location of the critical point in a region of the phase diagram accessible with energies available at the Super Proton Synchrotron accelerator~\cite{Grebieszkow:2009jr}. 

NA61/SHINE is a fixed target experiment located in the North Area of CERN. It studies hadron+hadron, hadron+ion and ion+ion collisions in the beam momentum range of 13-158(400)$A$ GeV/$c$. The detector is based on five Time Projection Chambers providing acceptance in the full forward hemisphere of the center of mass, down to $p_{T}=0$. Time of Flight detectors at the end of the apparatus provide additional particle identification. Centrality selection of ion+ion interactions is based on the measurement of the forward energy which does not bias fluctuation studies crucial in the search of the critical point. 
\section{Study of the onset of deconfinement}
\begin{figure}[htb]
\centerline{%
\includegraphics[height=4cm]{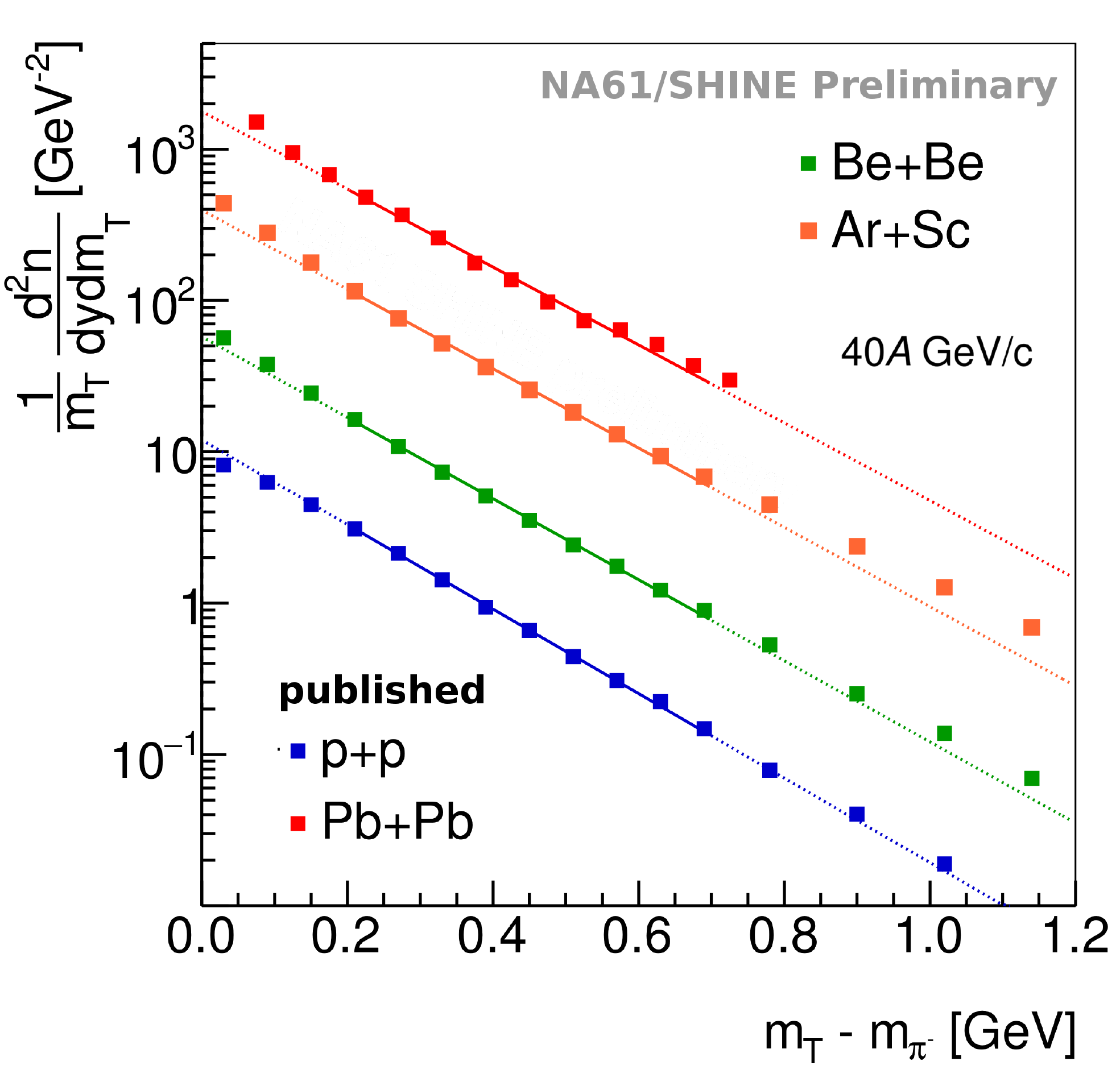}\includegraphics[height=4cm]{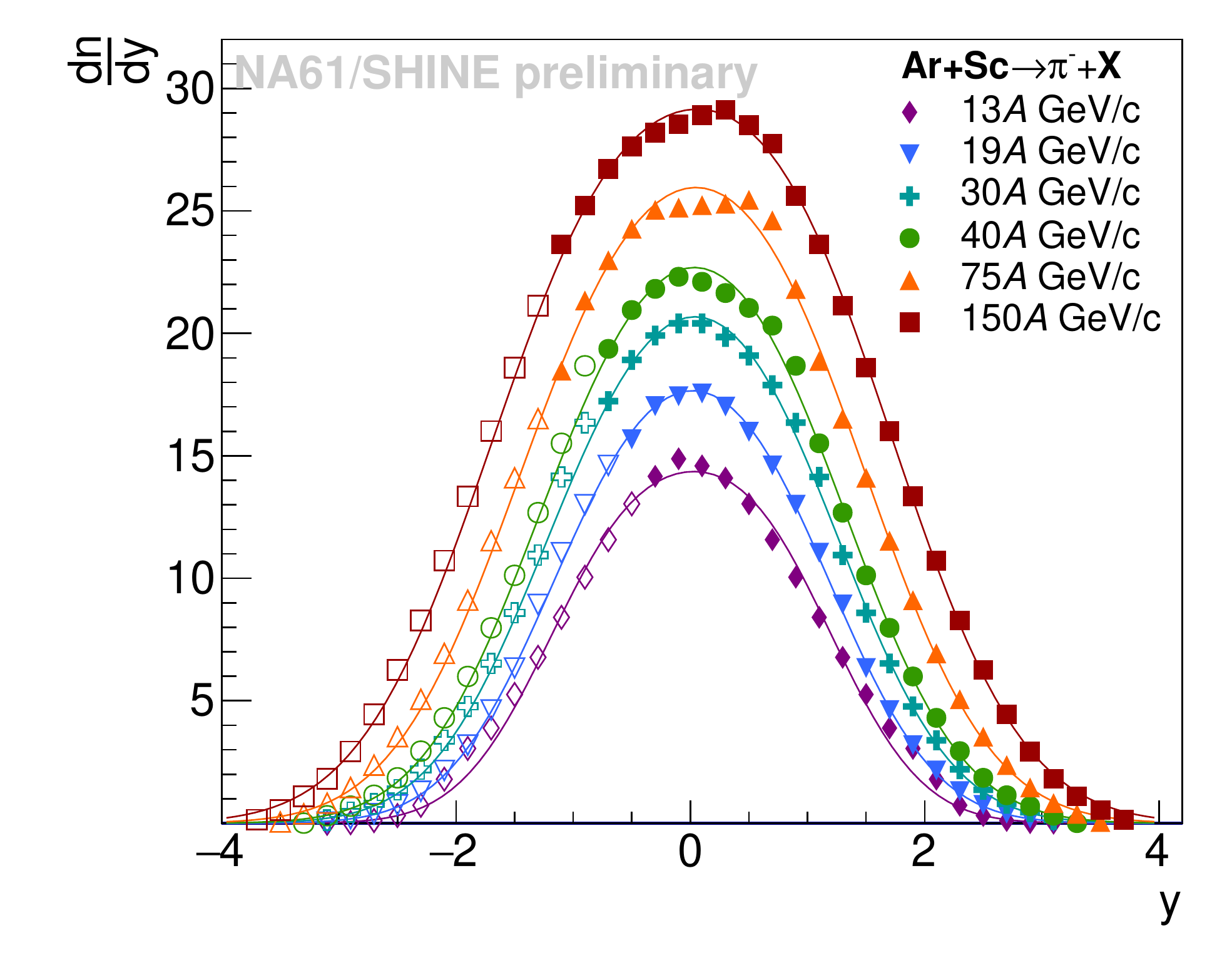}}
\caption{Left: Transverse mass spectra at mid-rapidity. Right: Rapidity spectra in Ar+Sc collisions at six beam momenta. A sum of two symmetrically displaced normal distribution of independent amplitudes was fitted to the distributions.}
\label{Fig:PiMSpectra}
\end{figure}
The spectra of $\pi^{-}$ were obtained using the so-called $h^{-}$ analysis method assuming that the majority of negatively charge particles are $\pi^{-}$ mesons. The contribution $(\approx10\%$) of other particles ($K^{-}$, $\bar{p}$) was estimated via the EPOS 1.99 model and subtracted. Transverse mass spectra at mid-rapidity at 40 GeV/$c$ in p+p, Be+Be, Ar+Sc and Pb+Pb (NA49)~\cite{Afanasiev:2002mx} interactions are shown in Fig.~\ref{Fig:PiMSpectra} (left). The $m_{T}$ spectra are exponential in p+p interactions, but deviate from this shape for heavier systems. Nevertheless, exponential functions were fitted in bins of rapidity to all systems in the $m_{T}$ range 0.2 - 0.7 GeV in order to extrapolate to the unmeasured region. Examples of the obtained rapidity spectra in Ar+Sc collisions at six studied beam momenta are shown in Fig.~\ref{Fig:PiMSpectra} (right). A sum of two symmetrically displaced normal distribution of independent amplitudes was fitted to spectra to calculate total yields. 
\begin{figure}[htb]
\centerline{%
\includegraphics[height=4cm]{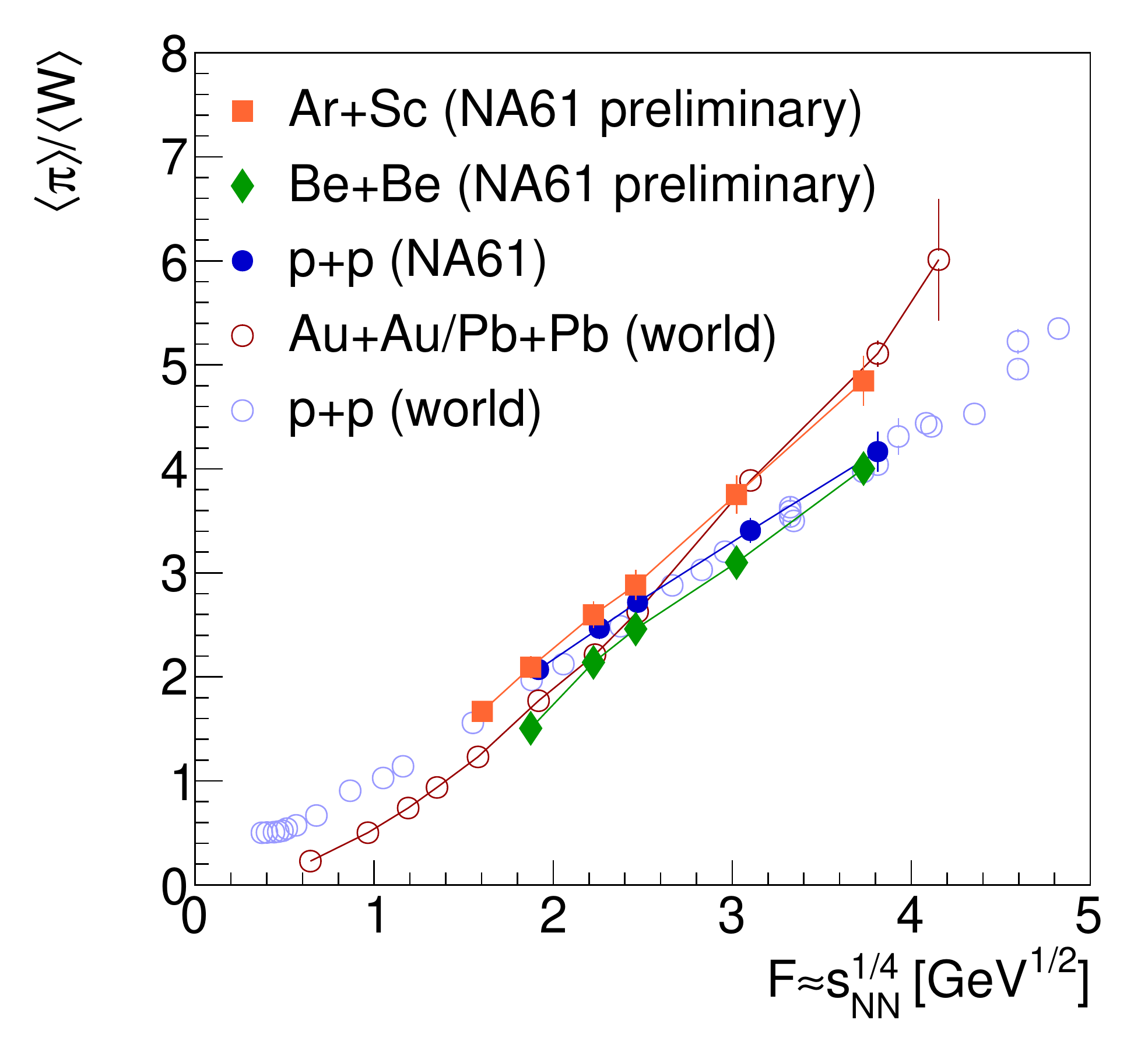}}
\caption{Energy dependence of the total pion multiplicity $\langle \pi \rangle$  calculated based on $\pi^{-}$ measurements. The data was divided by W the number of wounded nucleons in the reaction.}
\label{Fig:Kink}
\end{figure}
\begin{figure}[htb]
\centerline{%
    \includegraphics[height=4cm,trim={0.5cm 0 0 0},clip]{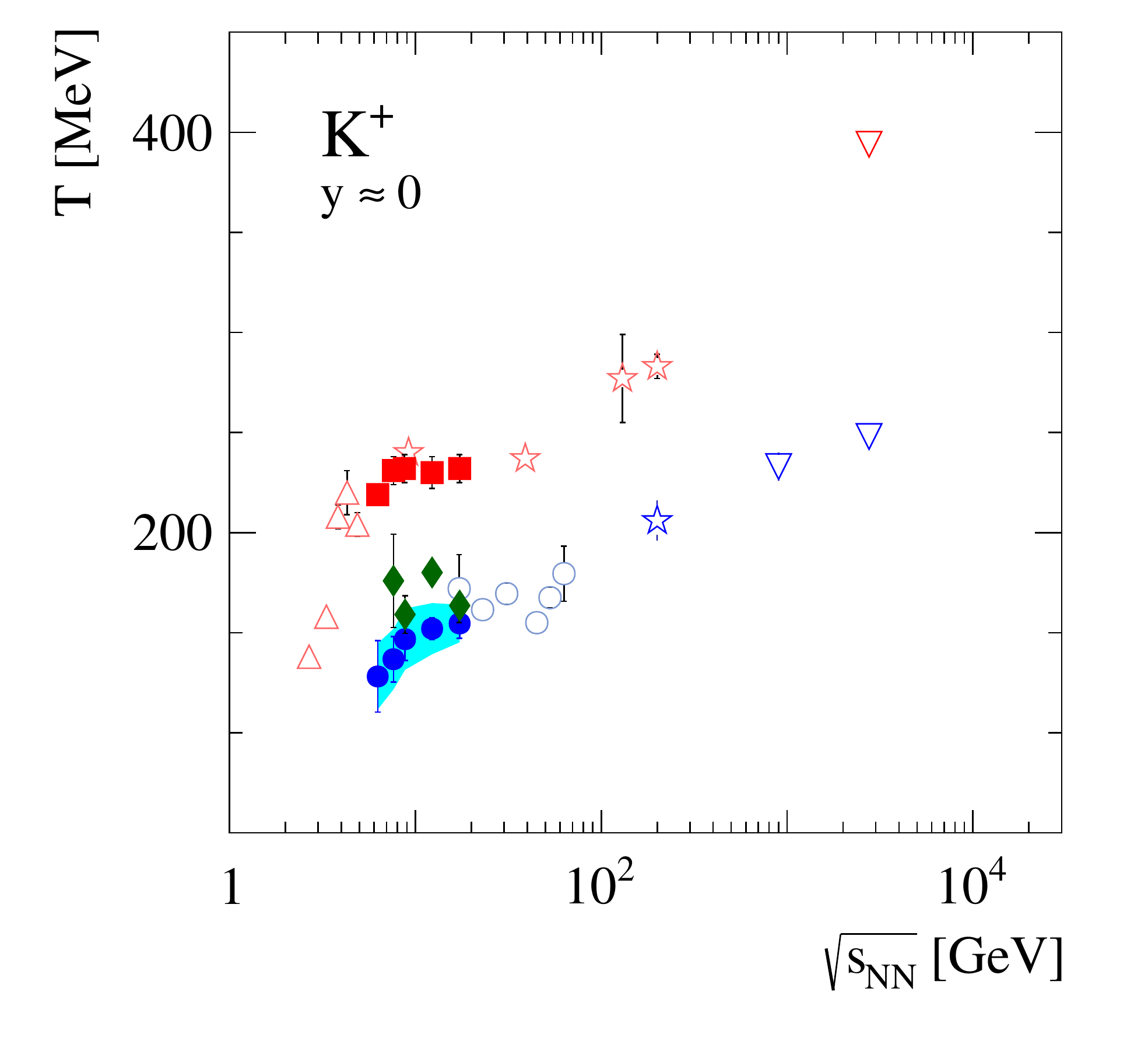}
    \includegraphics[height=4cm,trim={1.7cm 0 8.1cm 0},clip]{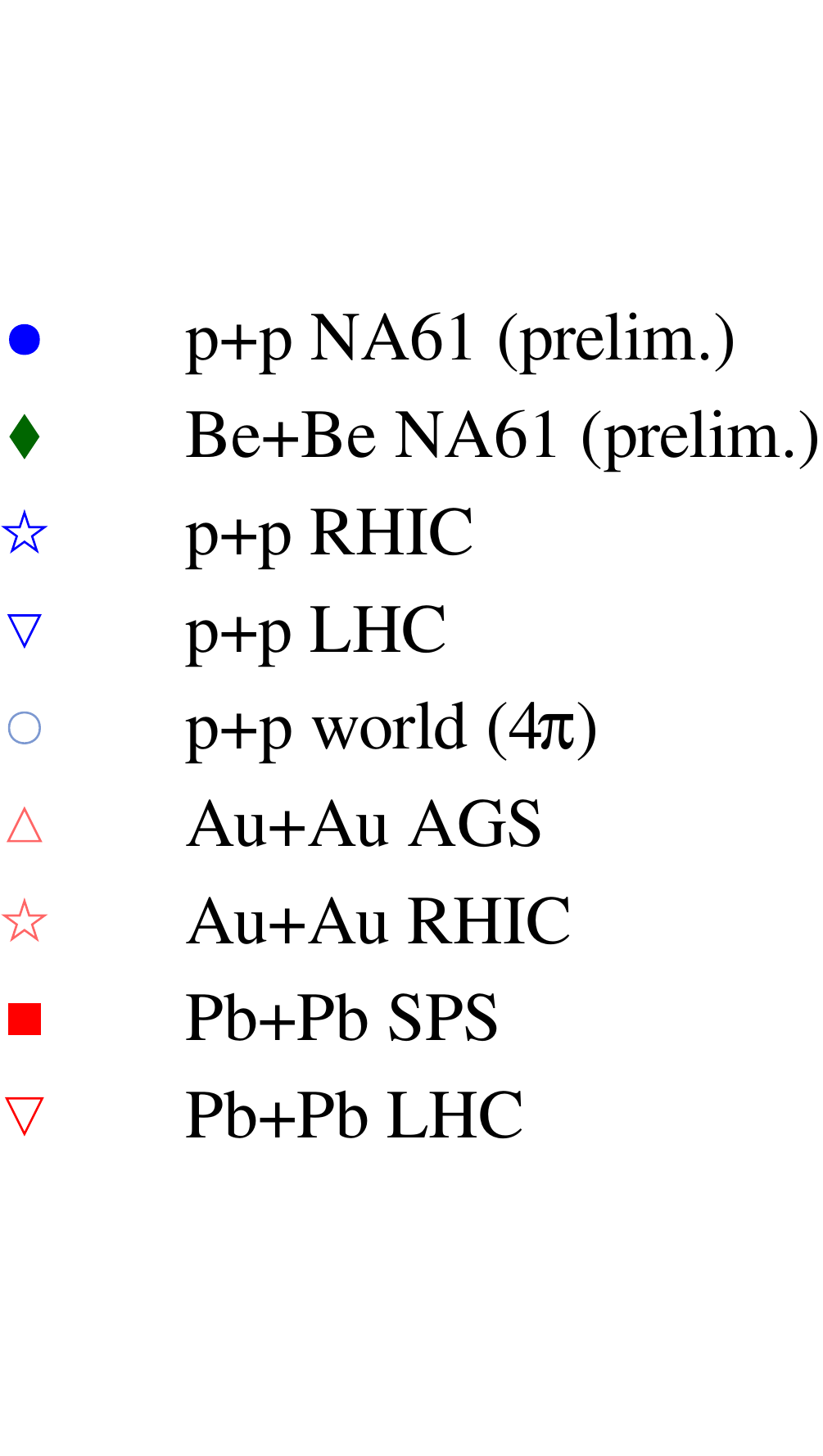}
    \includegraphics[height=4cm,trim={0.5cm 0 0 0},clip]{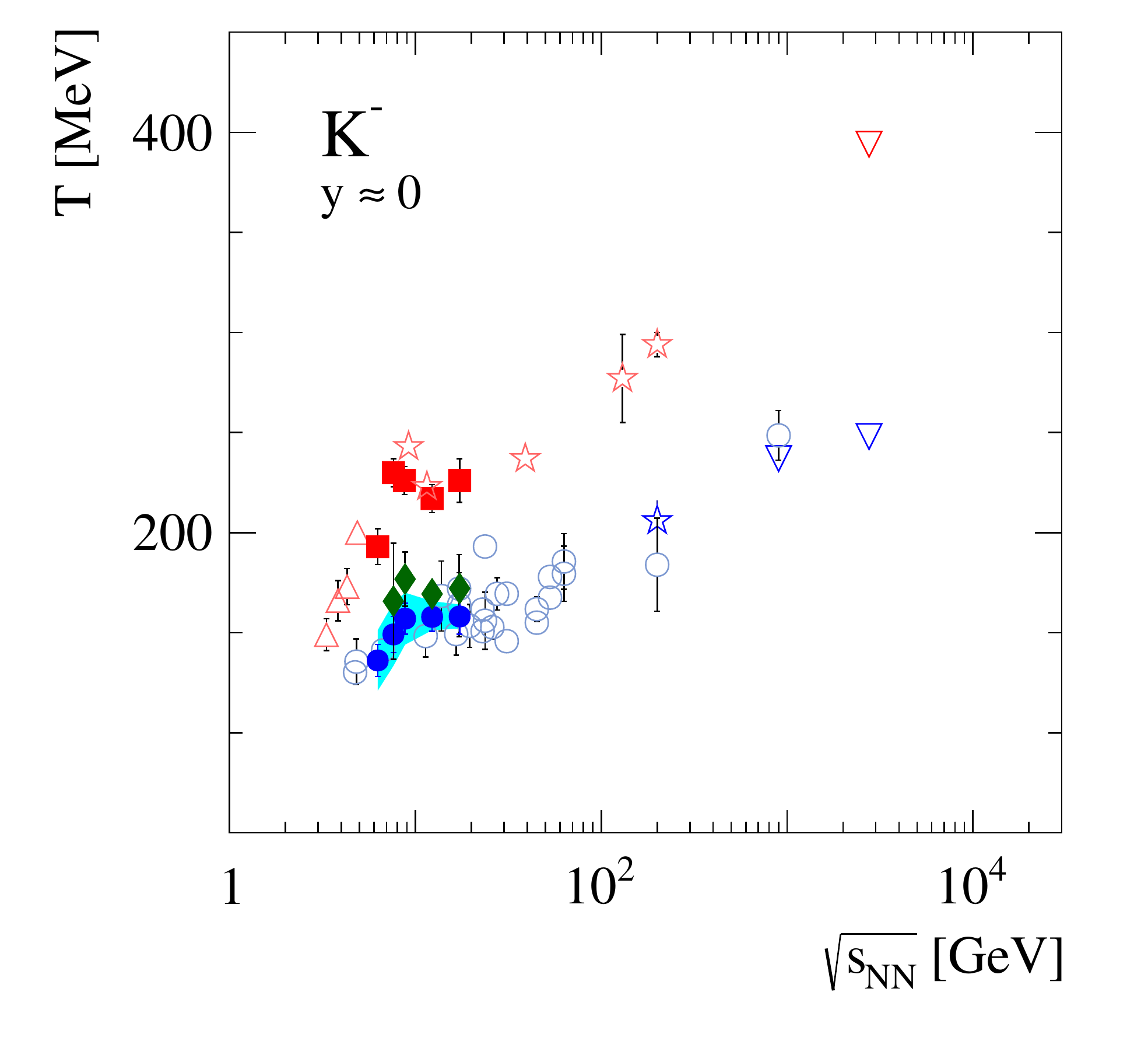}}
\caption{Energy dependence of the inverse slope parameter of $m_{T}$ spectra at mid-rapidity of $K^{+}$ (left) and $K^{-}$ (right).  The NA61/SHINE results on p+p interactions (full blue circles) and new results on Be+Be (full green diamonds) collisions are compared with world data on p+p and heavy ions (Pb+Pb and Au+Au). For details see Ref.~\cite{Aduszkiewicz:2017mei} and references therein.}
\label{Fig:Step}
\end{figure}

\begin{figure}[htb]
\centerline{%
    \includegraphics[height=4cm,trim={0.5cm 0 0 0},clip]{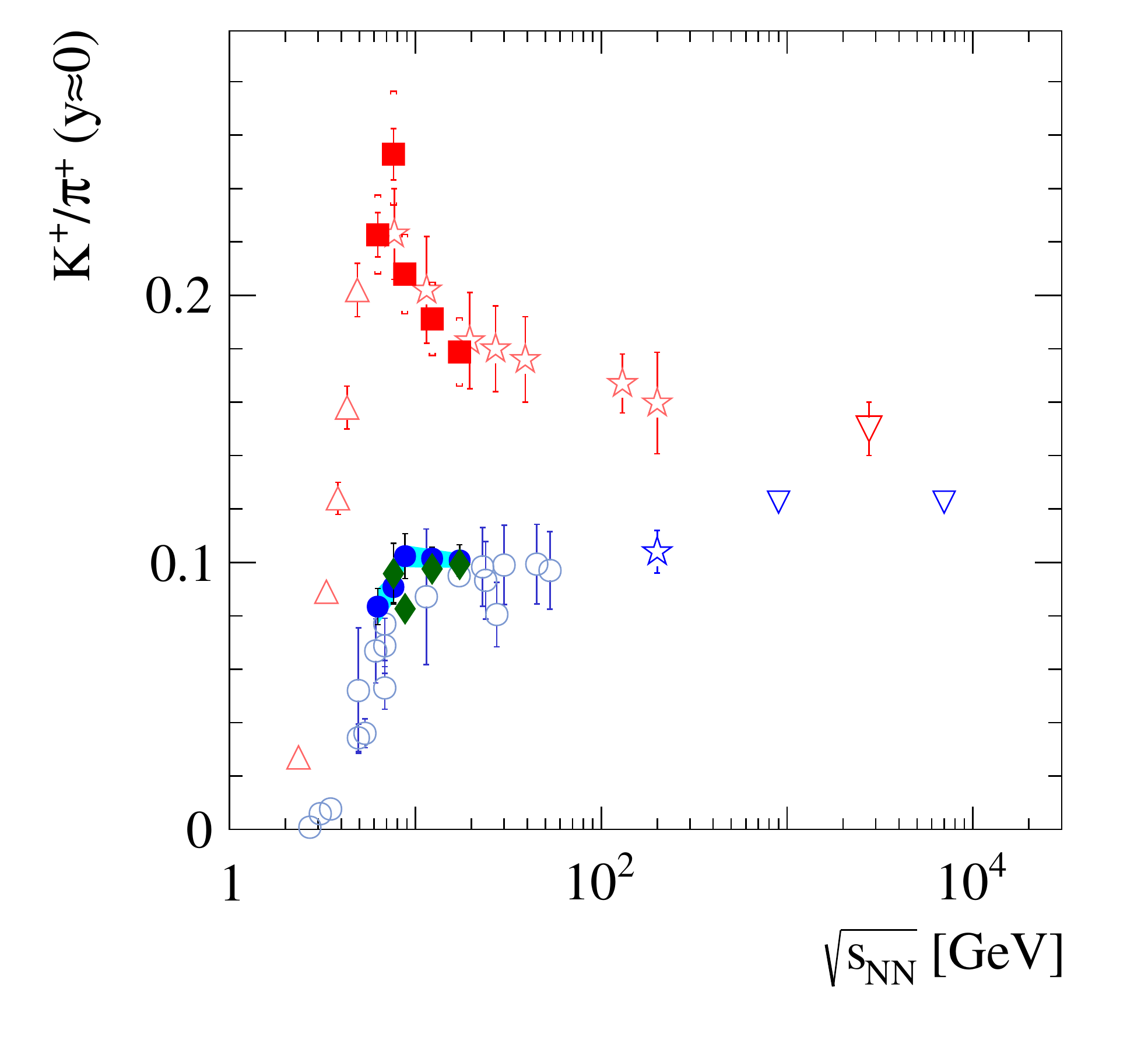}
    \includegraphics[height=4cm,trim={1.7cm 0 8.1cm 0},clip]{FIGs/legenda}
    \includegraphics[height=4cm,trim={0.5cm 0 0 0},clip]{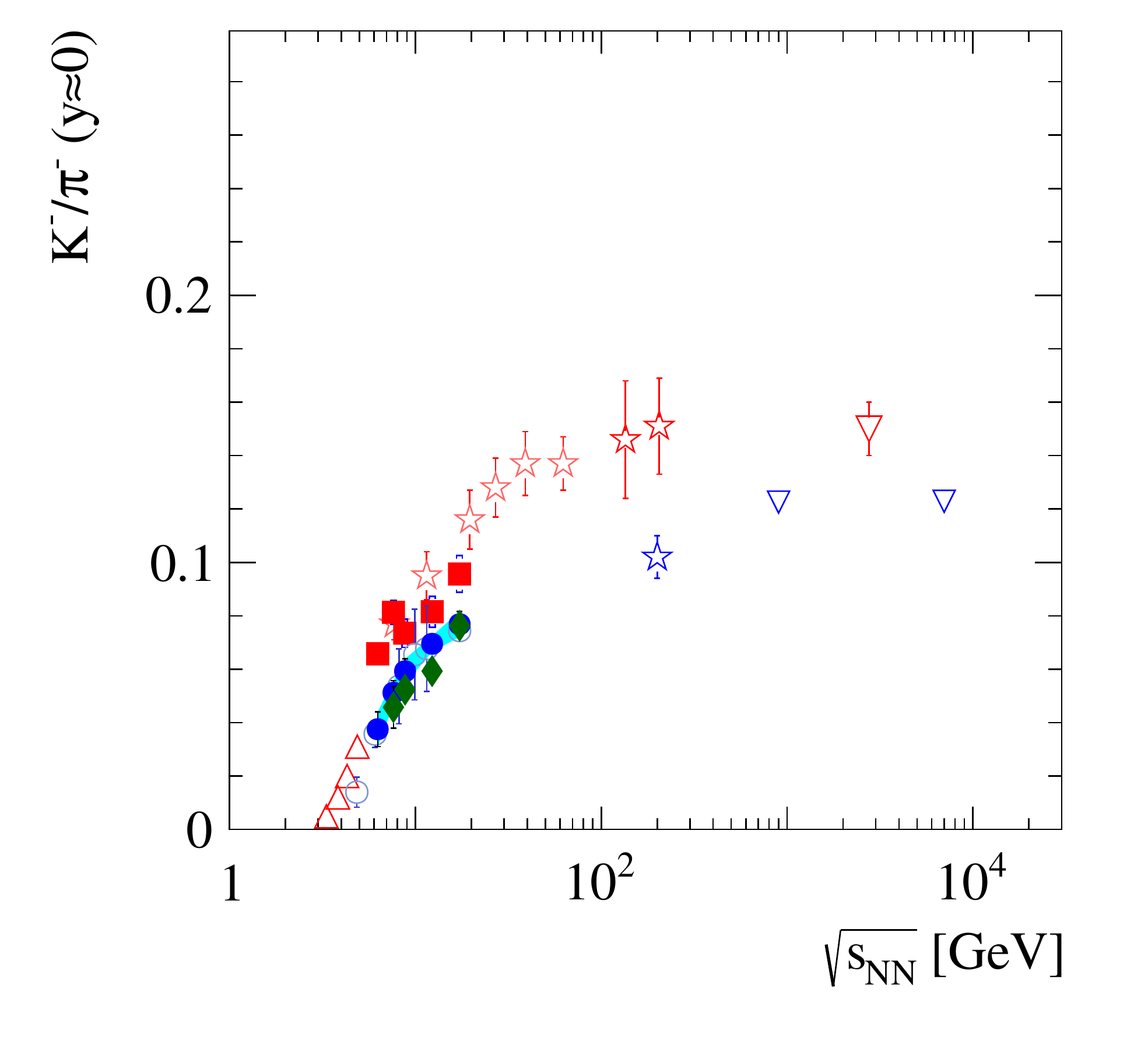}}
\caption{Energy dependence of the kaon to pion yield ratio at mid-rapidity for positively (left) and negatively (right) charged particles.}
\label{Fig:Horn}
\end{figure}

Signatures of the onset of deconfinement predicted by the Statistical Model of Early Stage (SMES): kink, step and horn~\cite{Gazdzicki:1998vd} are shown in Figs.~\ref{Fig:Kink},~\ref{Fig:Step},~\ref{Fig:Horn}. Their energy dependence for heavier systems differs from that in p+p and Be+Be interactions. The NA61/SHINE results on p+p interactions greatly improve the quality of the available data. They reveal rapid changes of the energy dependence in the SPS energy range suggesting that some properties of hadron production previously attributed to onset of deconfinement in heavy ion collisions my alo be present in p+p interactions~\cite{Poberezhnyuk:2015wea}.
\begin{figure}[htb]
\centerline{%
\includegraphics[height=4cm]{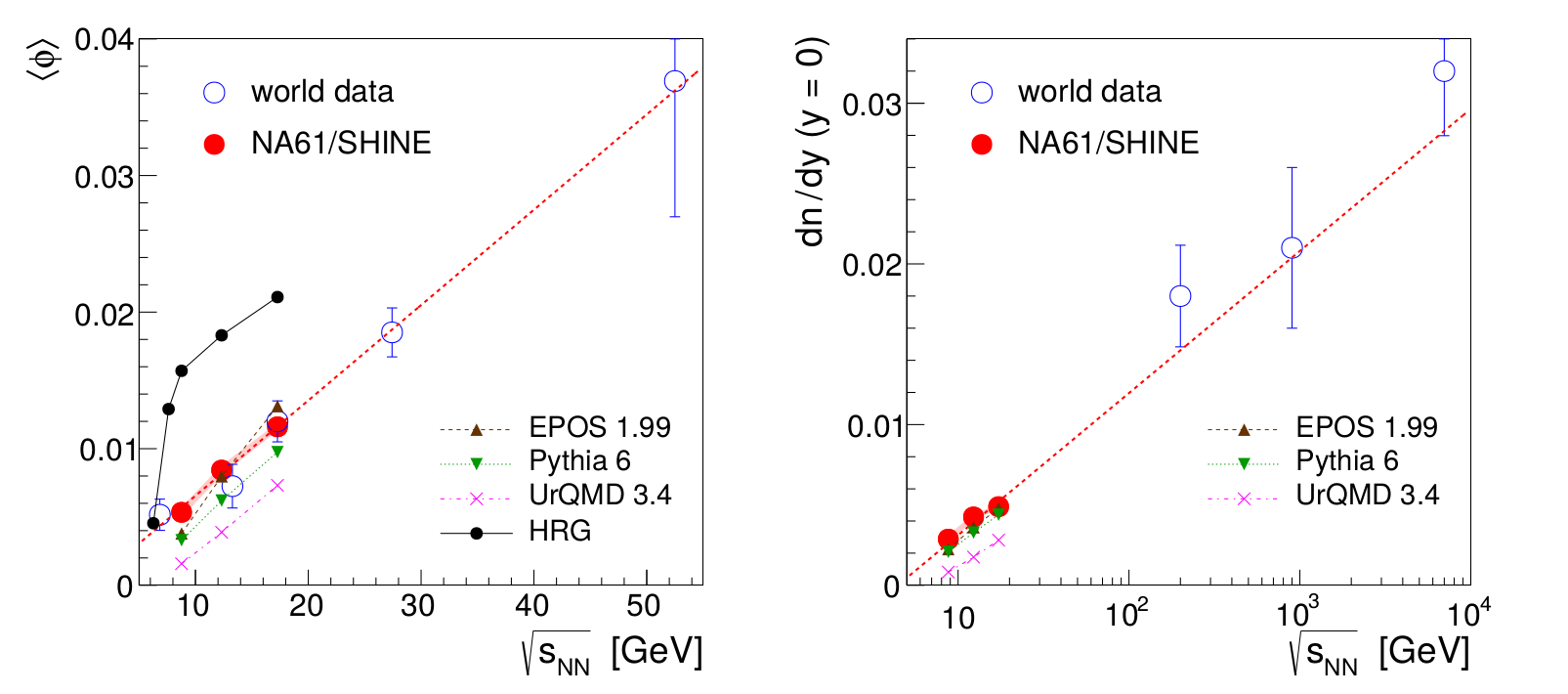}}
\caption{Energy dependence of $\phi$ production in p+p interactions: total yields (left) and mid-rapidity yield (right) }
\label{Fig:PhiMC}
\end{figure}
\begin{figure}[htb]
\centerline{%
\includegraphics[height=4cm]{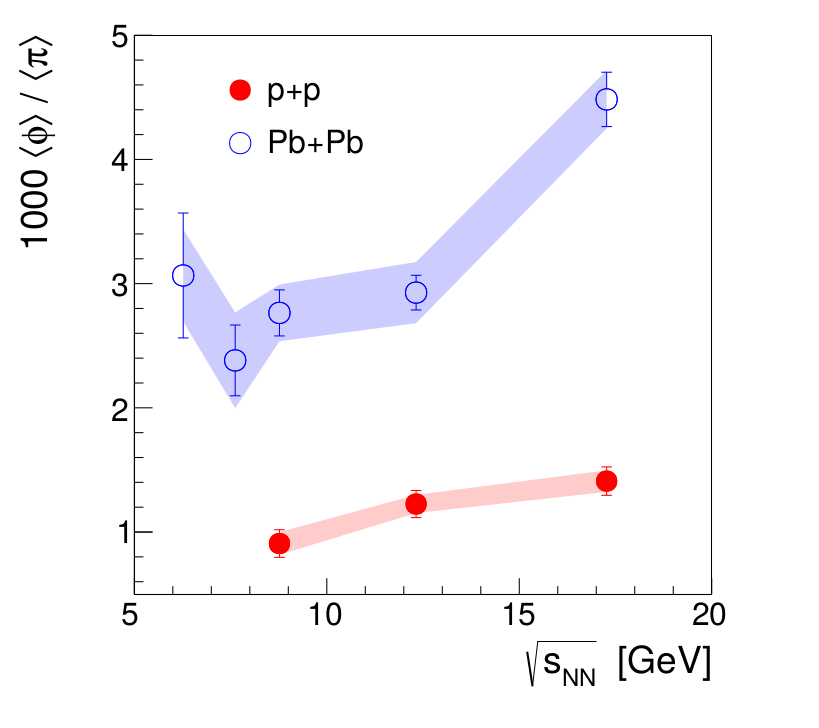}}
\caption{Energy dependence of $\phi$ to $\pi$ multiplicity ratio in p+p and Pb+Pb~\cite{Alt:2008iv} interactions.}
\label{Fig:PhiPb}
\end{figure}
NA61/SHINE performs also reference studies for heavier particles. Figures~\ref{Fig:PhiMC} and~\ref{Fig:PhiPb} show energy dependence of $\phi$ production.  The $\phi$ multiplicity is plotted in Fig.~\ref{Fig:PhiMC}~(left) and compared to various model predictions. Figure~\ref{Fig:PhiMC} (right) depicts the mid-rapidity yields. The best agreement with the data is shown by the EPOS 1.99 model although the energy dependence is too steep. Figure~\ref{Fig:PhiPb} demonstrates that the $\Phi$ to $\pi$ ratio is strongly enhanced in central Pb+Pb~\cite{Alt:2008iv} in comparison to p+p interactions. 
\section{Search for the critical point}
\begin{figure}[htb]
\centerline{%
\includegraphics[height=4cm]{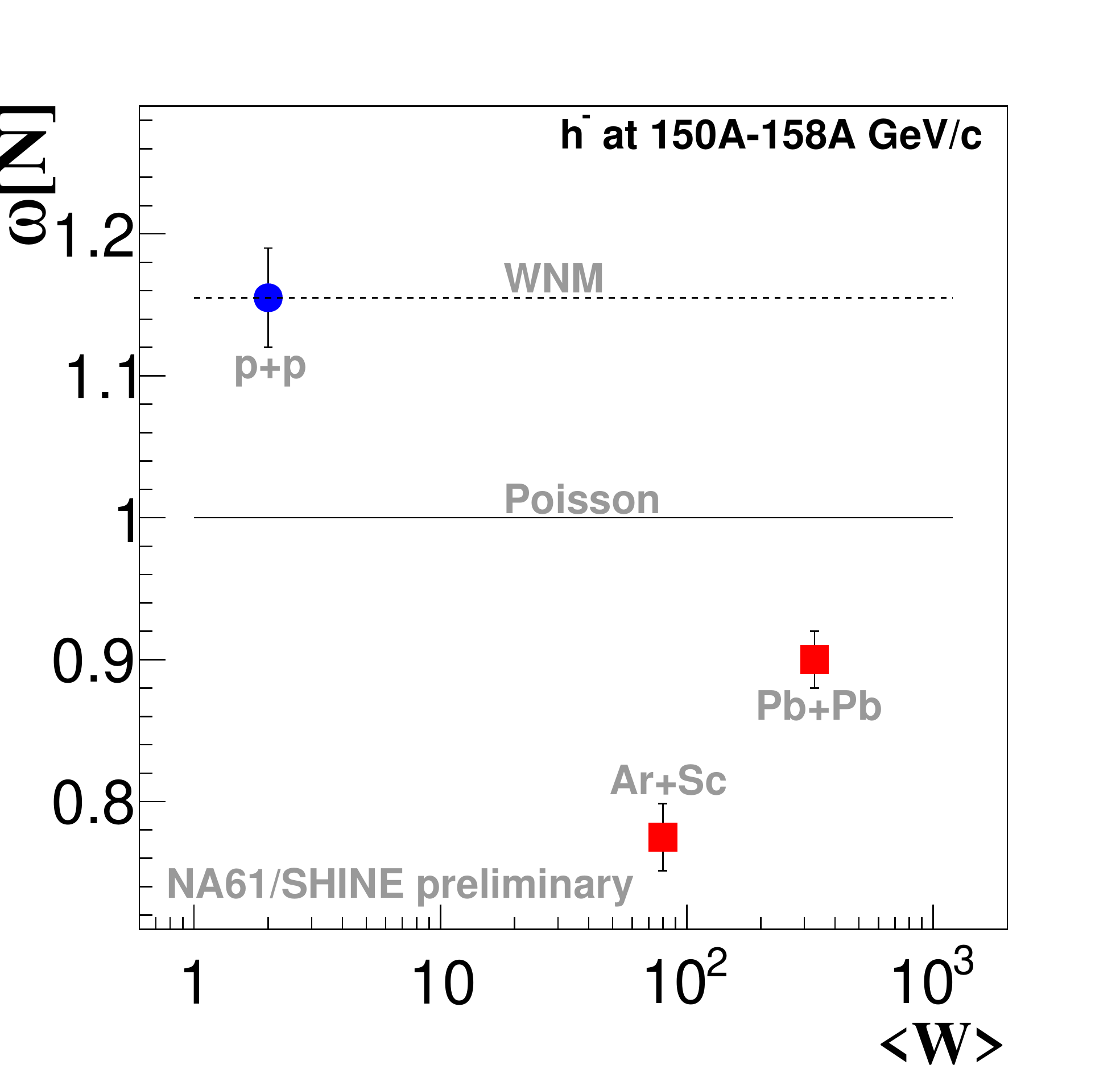}
\includegraphics[height=4cm]{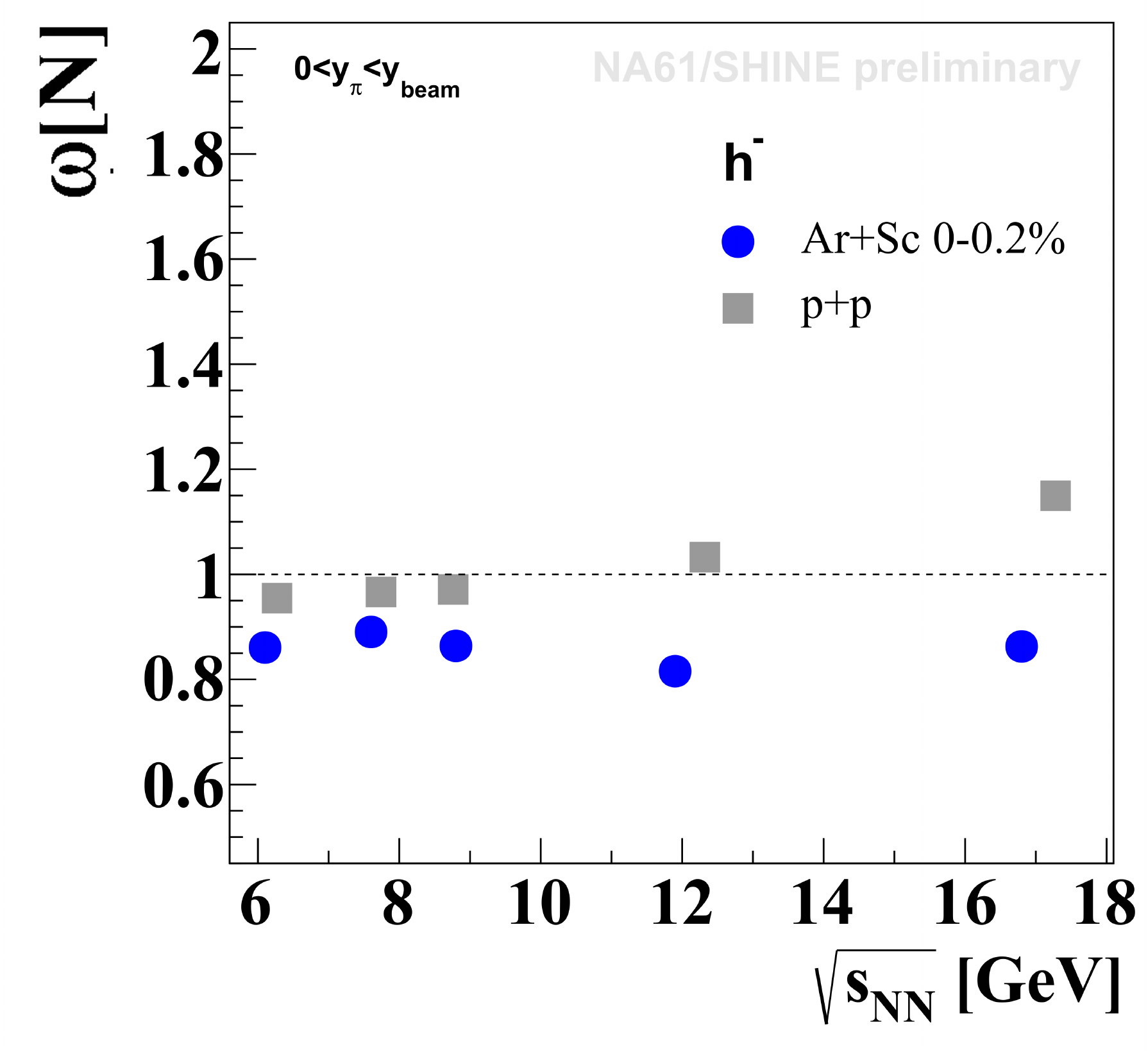}}
\caption{Scaled variance $\omega[N]$ of the negatively charged hadrons multiplicity distribution for p+p and 0.2$\%$ most central Ar+Sc collisions
measured  by  NA61/SHINE  and  1$\%$  most  central  Pb+Pb  collisions  measured  by  NA49~\cite{Grebieszkow:2009jr}. Left: System  size  dependence  of  results
calculated in the NA49 acceptance (for details see Ref.~\cite{Aduszkiewicz:2015jna}).  Wounded Nucleon Model~\cite{Bialas:1976ed} prediction of the minimal $\omega[N]$ value is marked with the dashed line. Value of $\omega[N]=1$ for the Poisson distribution is also shown. Right: Energy dependence of $\omega[N]$ calculated in the wider NA61/SHINE acceptance.}
\label{Fig:Omega}
\end{figure}
NA61/SHINE looks for indications of the critical point of strongly interacting matter by searching for non-monotonic dependence of fluctuations defined by moments higher than the first of distributions of event quantities. As fluctuations are sensitive to the system size and its fluctuations the following suitable families of quantities were used:
\begin{itemize}
	\item intensive quantities like $\omega=Var[N]/\langle N\rangle$, $S\sigma=\langle N^{3} \rangle_{c}/Var[N]$ and $\kappa\sigma^{2}=\langle N^{4} \rangle_{c}/Var[N]$ which are independent of system size but sensitive to its fluctuations ($\langle N^{3} \rangle_{c}$ and $\langle N^{4} \rangle_{c}$ are third and fourth cumulant)~\cite{Asakawa:2015ybt}.
	\item the strongly intensive quantity $\Sigma[P_{T},N]$, which measures fluctuations of transverse momentum and multiplicity independently of system size and its fluctuations~\cite{Gazdzicki:2013ana}. 
\end{itemize}
Fluctuations can not be corrected for limited acceptance thus comparison between different experiments or even systems is challenging. Results shown in a given figure were obtained in the same acceptance which may vary from one figure to another.
\begin{figure}[htb]
\centerline{%
\includegraphics[height=4cm]{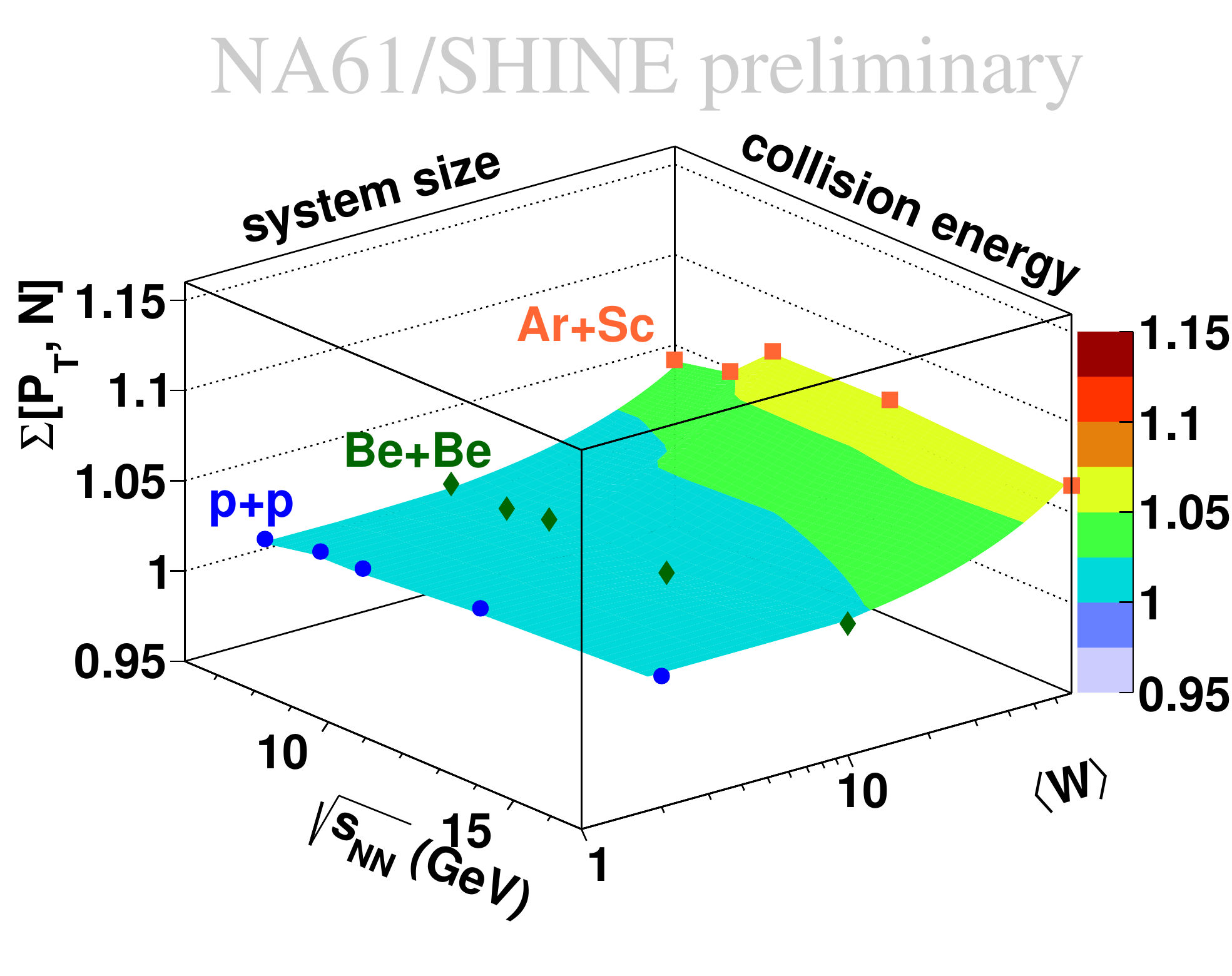}
\includegraphics[height=4cm]{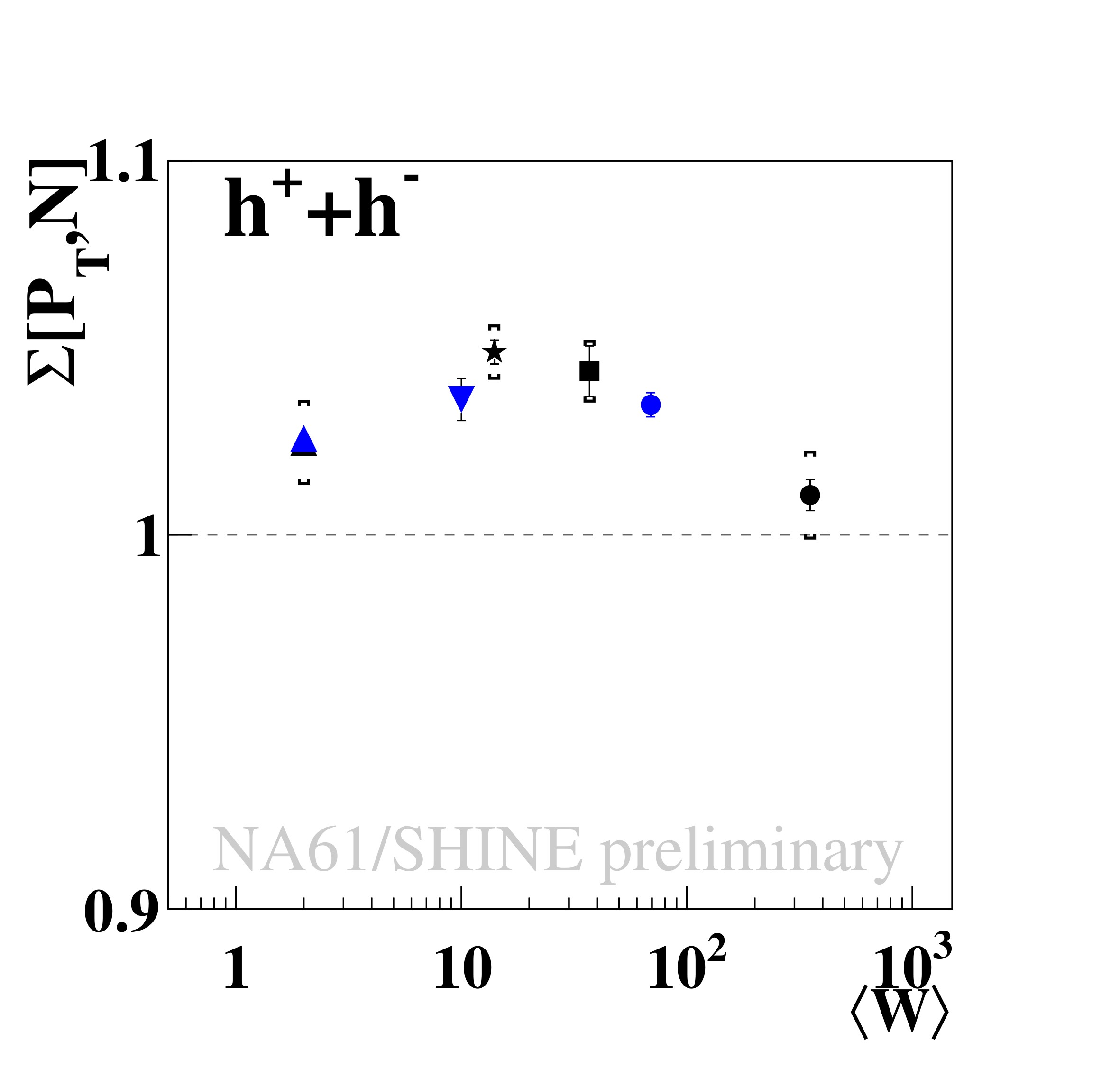}}
\caption{
Strongly intensive measure $\Sigma[P_{T},N]$ for negatively charged hadrons (left) and for all charged (right). Left plot shows results at five beam momenta for p+p, Be+Be and Ar+Sc in $0< y_{\pi}<y_{beam}$. Right plot shows NA61/SHINE p+p, Be+Be, Ar+Sc and NA49 p+p, C+C, Si+Si and Pb+Pb~\cite{Anticic:2015fla} at 150/158$A$ GeV/$c$ in $1.1 < y_{\pi} < 2.6$. }
\label{Fig:Sigma}
\end{figure}

Figure~\ref{Fig:Omega} (left) shows multiplicity fluctuations of negatively charged hadrons for p+p and central Ar+Sc and Pb+Pb~\cite{Grebieszkow:2009jr} collisions at 150/158$A$ GeV/$c$ beam momentum. Results contradict predictions of the Wounded Nucleon model (WNM) and ideal Boltzmann Grand Canonical Ensemble. Figure~\ref{Fig:Omega} (right) shows the energy dependence for p+p and Ar+Sc collisions. 
Two dimensional system size and energy dependence of $\Sigma[P_{T},N]$ is shown in Fig.~\ref{Fig:Sigma} (left). So far the results did not show non-monotonic dependence which could be attributed to the critical point. Figure~\ref{Fig:Sigma} (right) presents the system size dependence of $\Sigma[P_{T},N]$ measured by NA49 and NA61/SHINE at 150/158$A$ GeV/$c$. The results show consistent trends. 
\begin{figure}[htb]
\centerline{%
\includegraphics[height=3.1cm]{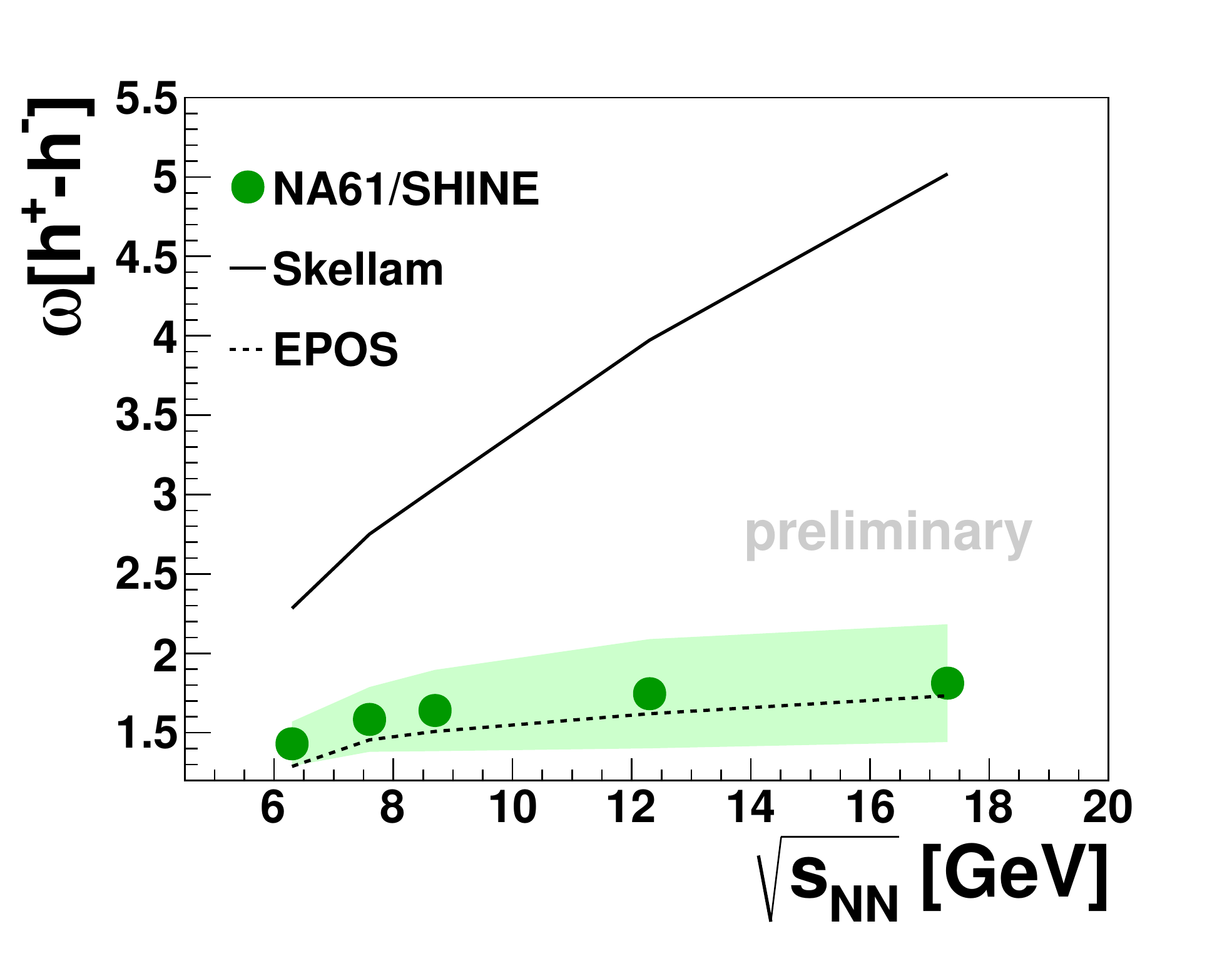}
\includegraphics[height=3.1cm]{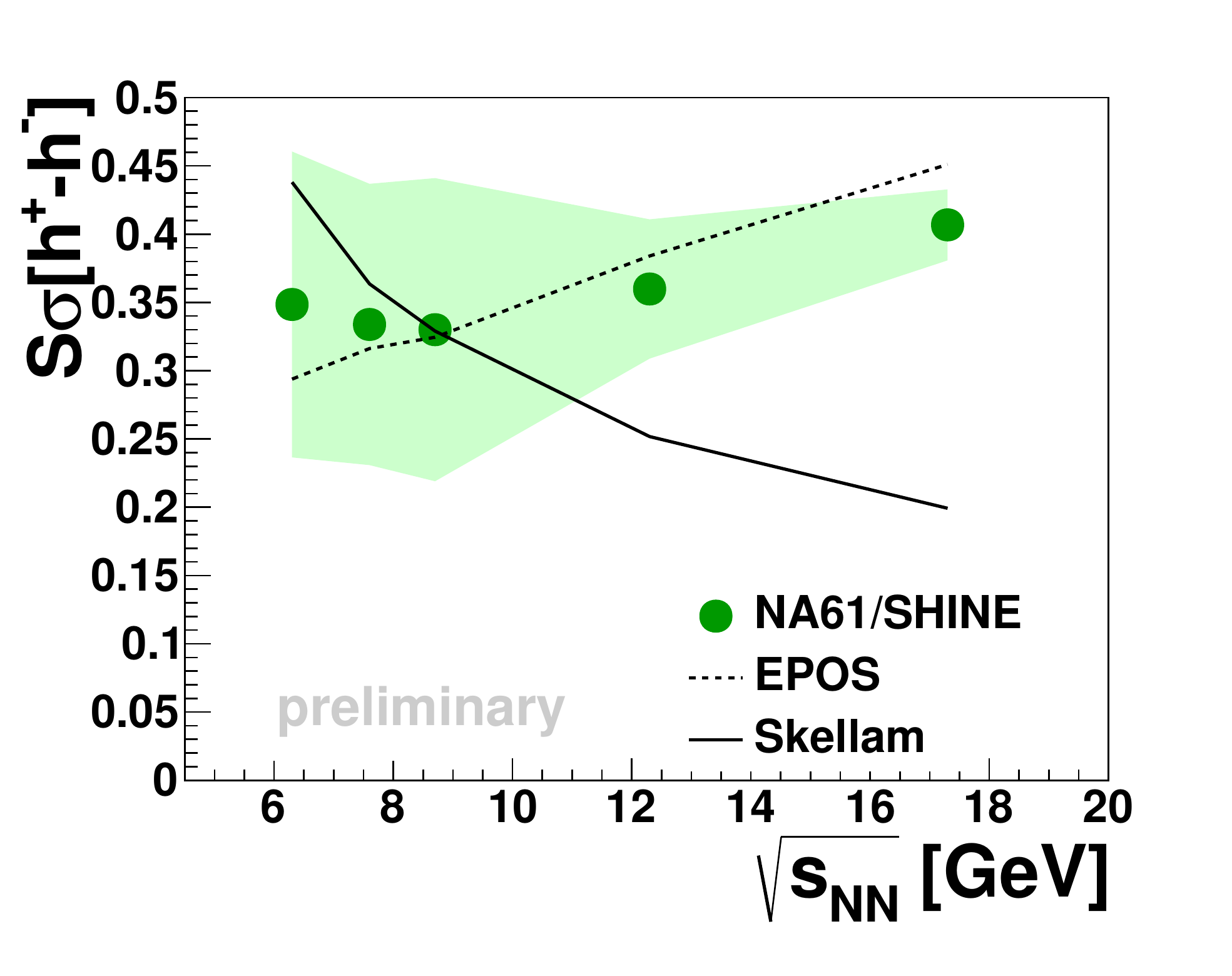}
\includegraphics[height=3.1cm]{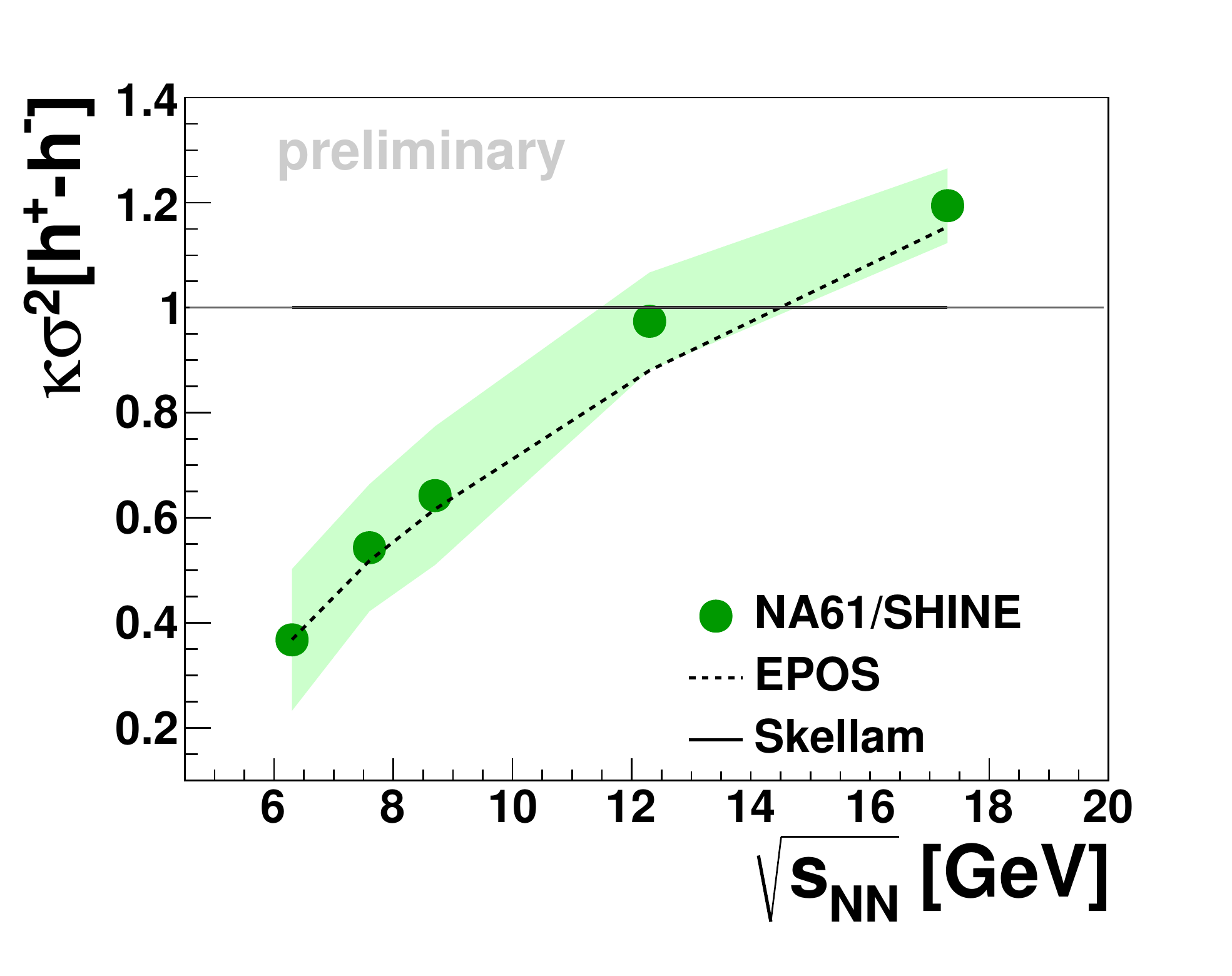}}
\caption{The energy dependence of $\omega[h^{+}-h^{-}]$, $S\sigma[h^{+}-h^{-}]$ and $\kappa\sigma^{2}[h^{+}-h^{-}]$ in p+p interactions.}
\label{Fig:HMNet}
\end{figure}

Higher order moments of distributions of event quantities are expected to be more sensitive to the critical point. Figure~\ref{Fig:HMNet} shows results on net-charge fluctuations measured by second, third and fourth order moments in p+p interactions. The obtained results do not agree with the Independent Particle Production model (Skellam distribution) but are described by the EPOS 1.99 model.
\section{Summary}
This contribution focused on recent results from the NA61/SHINE ion program aiming to study the onset of deconfinement and search for the critical point. Results on particle spectra and fluctuations were shown. Moreover, new $\phi$ production measurements in p+p interactions at 40, 80 and 158 GeV/$c$ beam momenta were reported. NA61/SHINE results on fluctuations quantified by second order moments do not show non-monotonic behaviour which could be connected to the critical point. Reference measurements of higher order moments in p+p interactions are available. The two-dimensional scan in energy and system size will be completed in 2018.  As an extension of this program NA61/SHINE plans to measure precisely open charm and multi-strange hyperon production in 2021$-$2024.

\small{{\it Acknowledgements:} This  work  was  partially  supported  by  the  National  Science
Centre, Poland grants 2016/21/D/ST2/01983 and 2015/18/M/ST2/00125.}


\end{document}